\documentclass[]{aastex631}
\usepackage{amssymb,amsmath}
\usepackage{graphicx}
\usepackage{xcolor,natbib}
\usepackage{multirow}
\usepackage[T1]{fontenc}
\usepackage{ae,aecompl}
\usepackage{newtxtext, newtxmath}
\usepackage{float}
\usepackage{subfigure}
\usepackage{longtable}
\usepackage{enumitem}
\usepackage{longtable,listings}
\usepackage[flushleft]{threeparttable}
\usepackage{parcolumns}
\def\oc3{[O~{\sc iii}]$_c$}
\def\ob3{[O~{\sc iii}]$_b$}

\submitjournal{ApJ}

\shorttitle{Negative AGN feedback?}
\shortauthors{Zhang XueGuang}

\begin{document}

\title{Different dependence of narrow H$\alpha$ line luminosity on optical continuum luminosity between 
starforming galaxies and Type-2 AGN: globally negative AGN feedback in local Type-2 AGN?}

\correspondingauthor{XueGuang Zhang}
\email{xgzhang@gxu.edu.cn}
\author{XueGuang Zhang$^{*}$}
\affiliation{Guangxi Key Laboratory for Relativistic Astrophysics, School of Physical Science and Technology, 
GuangXi University, Nanning, 530004, P. R. China}

\begin{abstract} 
	In this manuscript, clues are provided to support globally negative AGN feedback on star formation in 
the host galaxies of the local low-redshift SDSS Type-2 AGN, based on the different dependence of narrow 
H$\alpha$ line luminosity $L_{H\alpha}$ on optical continuum luminosity $\lambda L_{cont}$ between the 
starforming galaxies and the Type-2 AGN. Through the measured $L_{H\alpha}$ and $\lambda L_{cont}$ in SDSS 
starforming galaxies, there is a strong linear correlation between $\lambda L_{cont}$ and $L_{H\alpha}$, 
accepted as a standard correlation without effects of AGN activity. Meanwhile, considering apparent contributions 
of AGN activity to narrow H$\alpha$ line emissions in the Type-2 AGN, the correlation between 
$\lambda L_{cont}$ and $L_{H\alpha}$ in the SDSS Type-2 AGN leads to statistically lower $L_{H\alpha}$ in the 
Type-2 AGN than in the starforming galaxies, with significance level higher than 5$\sigma$, even after considering 
necessary effects (including effects of host galaxy properties), leading to accepted conclusion on the globally 
negative AGN feedback in the local Type-2 AGN. Meanwhile, properties of Dn(4000) and H$\delta_A$ can provide 
indirect clues to support the globally negative AGN feedback in the local Type-2 AGN, due to older stellar ages 
in the Type-2 AGN. Moreover, it is interesting to expect more than 50\% narrow H$\alpha$ emissions globally 
suppressed in the host galaxies of the Type-2 AGN relative to the starforming galaxies. The results not only 
support globally negative AGN feedback in the local Type-2 AGN, but also show further clues on the quantification 
of suppressions of star formation by the globally negative AGN feedback.
\end{abstract}


\section{Introduction}

	AGN feedback through galactic-scale outflows plays a key role in galaxy evolution, leading to the 
tight connections between Active Galactic Nuclei (AGN) and host galaxies \citep{ref1, ref2, gb12, ref3, ref4, 
ref5, ref6, ref7, bn18, rm19, kk20, rh20, ch21, kr21, ss21, pb22}. Both observational and theoretical results 
have shown clear impacts of either positive or negative AGN feedback on star formation in the host galaxies of 
AGN. \citet{ref9} have reported observational evidence to support negative AGN feedback in the nearest quasar 
Mrk 231, due to the detected giant molecular outflow with higher mass rate than the detected star formation 
rate in the host galaxy, therefore halting star formation. \citet{ref15} have shown evidence to support negative 
AGN feedback, because rapid star formations are common in AGN host galaxies but that the most vigorous star 
formations cannot be observed in AGN with X-ray luminosities larger than $10^{44}{\rm erg\cdot s^{-1}}$. 
\citet{ref11} have reported evidence for the negative AGN feedback, based on the AGN with strong outflow 
signatures being hosted in galaxies that are more quenched than galaxies with weaker outflow signatures. 
\citet{ref8} have also shown evidence to support the negative AGN feedback, based on inside-out quenching of 
star formations in radio-mode AGN host galaxies which have older stellar populations through a sample of 406 
AGN subdivided into radio-quiet and radio-mode AGN. Meanwhile, evidence to support positive AGN feedback can 
be found in the literature. \citet{ref13} have discussed the positive AGN feedback in gas-rich phases by 
over-compressing cold dense gas. \citet{ref10} have shown the positive AGN feedback, because much higher 
star formation rates in the AGN with pronounced radio jets than in the purely X-ray-selected ones. \citet{ref12} 
have reported positive feedback scenario in NGC 5728, due to higher star formation rates in the encountering 
region where the ionized gas outflows encounter the star formation ring at 1kpc radius. \citet{mp22} have 
shown that there are no apparent signs of negative AGN feedback, after comparing host galaxy properties of 
far-infrared AGN and non-AGN green valley galaxies. Therefore, it is interesting to provide further clues 
on the contradictory effects of AGN feedback on star formations, through different methods, which is the 
main objective of this manuscript.

	Optical spectroscopic properties of starforming galaxies with no contribution of central AGN activity 
can be well applied to trace starforming histories, leading to the strong dependence of the star formation 
rates (SFRs) on the narrow H$\alpha$ line luminosities ($L_{H\alpha}$) in galaxies as well discussed in 
\citet{ref16, ref17, ref18, ken94, pw07, ref19, ll11, vb21}: 
\begin{equation}
	SFRs~\propto~L_{H\alpha}^{\beta}
\end{equation}. 
In this manuscript, among all the narrow emission-line main galaxies in SDSS DR16 (Sloan Digital Sky Survey, 
Data Release 16) \citep{ref31}, starforming galaxies with high-quality narrow emission lines can be collected, 
based on the dividing line applied in the BPT diagram \citep{bpt81, ka03a, ke06, km13, ref20, ref21, ref22, 
zh22a} through the narrow emission line flux ratios of [O~{\sc iii}]$\lambda5007$\AA~ to H$\beta$ (O3HB) and 
of [N~{\sc ii}]$\lambda6583$\AA~ to H$\alpha$ (N2HA), which will be discussed in Section 2. Based on properties 
of continuum emissions and narrow H$\alpha$ emissions of the starforming galaxies with no contribution of AGN 
activity on the spectroscopic features, dependence of narrow H$\alpha$ line luminosity on optical continuum 
luminosity $\lambda L_{cont}$ in the starforming galaxies can be used as a standard candle to check effects 
of AGN feedback in Type-2 AGN.

	Not similar as the starforming galaxies, Type-2 AGN have optical spectroscopic narrow emission lines 
including apparent contributions of central AGN activity. Based on commonly accepted and constantly being 
improved unified model \citep{ref25, ref26} of AGN, emissions from central accretion disks and from central 
broad emission line regions are totally obscured by central dust torus \citep{db15, nl16, ma17, zhu18, pn21, 
zh22b, zh23} in the Type-2 AGN. Therefore, narrow emission lines of Type-2 AGN have contributions from both 
central AGN activity and starforming, however, both the continuum emissions and the narrow absorption features 
in host galaxies of the Type-2 AGN have few contributions of central AGN activity, which are strongly supported 
by optical spectra (with emission line features being masked out) of Type-2 AGN described by pure stellar 
templates without considerations of AGN contributions as discussed in Section 3. Hence, studying properties 
of the narrow H$\alpha$ emissions in a large sample of Type-2 AGN can provide clues on effects of AGN feedback, 
after considering effects of AGN activity on the observational narrow H$\alpha$ line emissions but few effects 
on the observational continuum emissions in the host galaxies of Type-2 AGN.

	The manuscript is organized as follows. Section 2 presents the data samples of starforming galaxies and 
Type-2 AGN. Section 3 presents our main results and necessary discussions. Section 4 gives our summary and 
conclusions. And in this manuscript, we have adopted the cosmological parameters of 
$H_{0}=70{\rm km\cdot s}^{-1}{\rm Mpc}^{-1}$, $\Omega_{\Lambda}=0.7$ and $\Omega_{\rm m}=0.3$.

\begin{figure*}
\centering\includegraphics[width = 18cm,height=9cm]{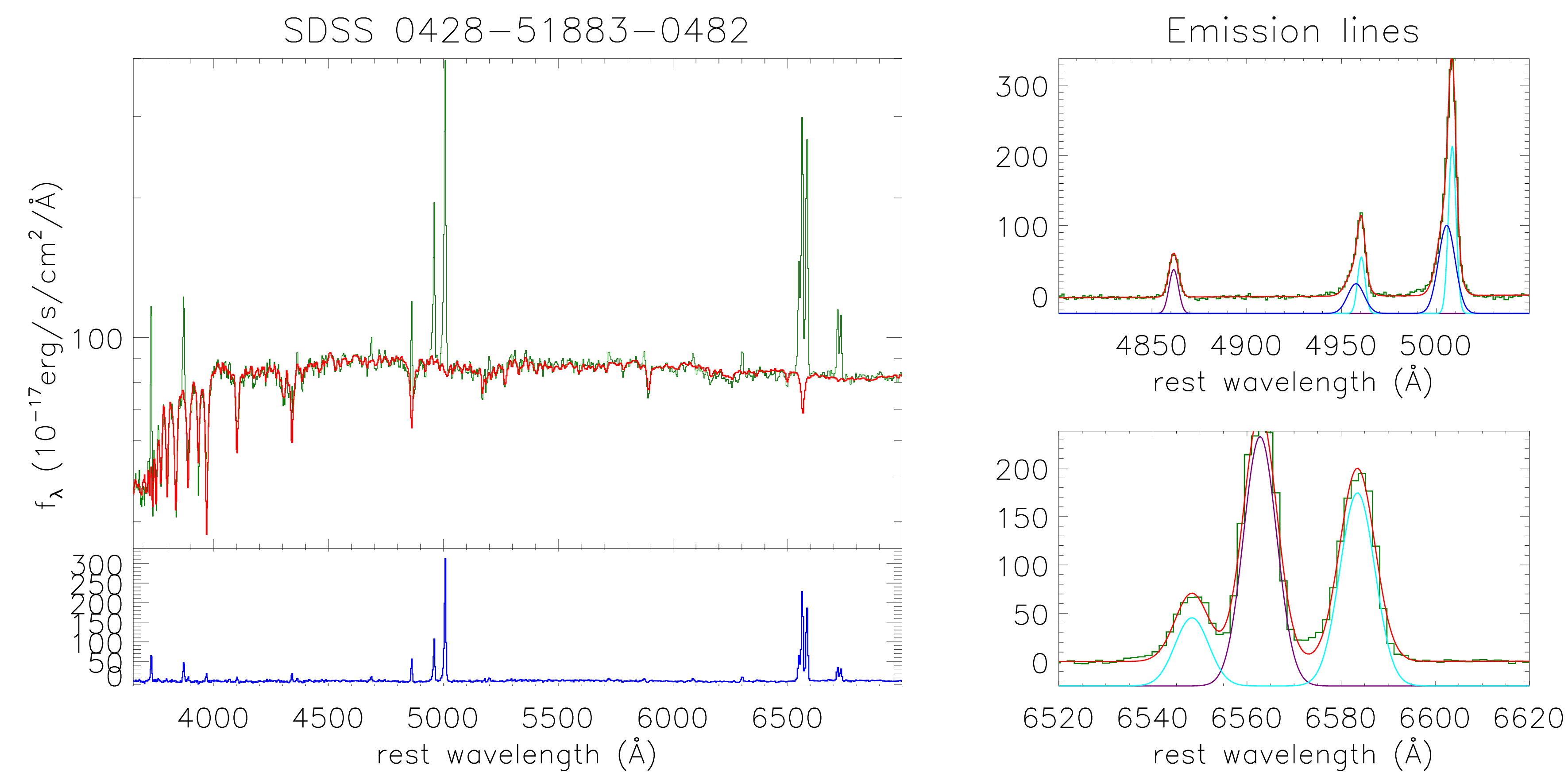}
\caption{Left panels show examples on the SSP method determined stellar continuum (in top left panel) and the 
corresponding line spectrum (in bottom left panel) in the Type-2 AGN SDSS 0533-51994-0031. Right panels show 
the corresponding best fitting results to the emission lines, after subtractions of the determined stellar 
continuum. In the left panels, solid lines in dark green, red and blue represent the observed SDSS spectrum, 
the SSP method determined stellar continuum and the line spectrum after subtractions of the stellar continuum, 
respectively. In right panels, solid dark green line shows the line spectrum, solid red line shows the 
best-fitting results. In the top right panel, solid purple line shows the determined narrow H$\beta$, solid 
cyan lines and solid blue lines show the determined core and shifted-wing related broad components of 
[O~{\sc iii}] doublet, respectively. In the bottom right panel, solid purple line and solid cyan lines show 
the determined narrow H$\alpha$ and [N~{\sc ii}] doublet, respectively. In the top left panel, in order to 
show clearer absorption features, the Y-axis is shown in logarithmic coordinate.}
\label{ssp}
\end{figure*}

\section{Data Samples}

	All the low redshift ($z<0.35$) main galaxies with high-quality spectra (median spectral signal-to-noise 
ratio $S/N$ larger than 20) are firstly collected from SDSS DR16, through the SDSS provided SQL (Structured 
Query Language) Search tool (\url{http://skyserver.sdss.org/dr16/en/tools/search/sql.aspx}) by the following 
query
\begin{lstlisting}
SELECT S.plate, S.fiberid, S.mjd, S.z, S.snmedian,  
       P.petroR50_u, P.petroR50err_u, P.petroR50_g, P.petroR50err_g, 
       P.petroR50_r, P.petroR50err_r, P.petroR50_i, P.petroR50err_i, 
       P.petroR50_z, P.petroR50err_z, P.petroR90_u, P.petroR90err_u,
       P.petroR90_g, P.petroR90err_g, P.petroR90_r, P.petroR90err_r, 
       P.petroR90_i, P.petroR90err_i, P.petroR90_z, P.petroR90err_z, 
       P.devab_u, P.devab_g, P.devab_r, P.devab_i, P.devab_z, 
       M.mstellar_median, M.mstellar_err
FROM SpecObjall as S JOIN PhotoObjAll as P ON S.bestobjid = P.objid
     JOIN stellarMassPCAWiscM11 as M ON M.specobjid = S.specobjid
WHERE S.class='GALAXY' and S.z<0.35 and S.zwarning=0 and S.snmedian > 20 
\end{lstlisting}
Here, the restriction $z<0.35$ is applied, to ensure that the narrow H$\alpha$ and the [N~{\sc ii}] emission 
lines to be totally covered in the SDSS spectra, which will be used to do classifications of the main galaxies 
through the BPT diagrams. Meanwhile, properties of the inverse concentration parameter $IC$ will be discussed 
in Section 3, therefore, the SDSS public database of 'PhotoObjAll' is also considered in the SQL query, in 
order to collect the corresponding photometric information. The detailed descriptions on the database 
'PhotoObjAll' containing full photometric catalog quantities and on the database 'SpecObjall' containing all 
the spectroscopic information can be found in \url{https://skyserver.sdss.org/dr16/en/help/docs/tabledesc.aspx}. 
Furthermore, properties of the total stellar mass (the parameter mstellar\_median and the corresponding 
uncertainty mstellar\_err) will be discussed in Section 3, therefore, the SDSS public database of 
'stellarMassPCAWiscM11' as described in detail in \citet{ms11, ch12} is also considered in the SQL query.

	Before proceeding further, spectroscopic features of the collected main galaxies are carefully checked, 
in order to measure emission lines after subtractions of the SSP method determined stellar continuum. In this 
manuscript, the commonly applied SSP (Simple Stellar Population) \citep{ref33, kh03, cm05, cp13, lc16, ref32, 
wc19} method is accepted to determine contributions of stellar lights in the SDSS spectra, with the 39 SSP 
templates discussed in \citet{ref33, kh03} which include the population age from 5Myr to 12Gyr with three 
solar metallicities (Z~=~0.008,~0.05,~0.02). Through the Levenberg-Marquardt least-squares minimization method 
(the known MPFIT package, \url{https://pages.physics.wisc.edu/~craigm/idl/cmpfit.html})\citep{ma09}, sum of 
the strengthened, broadened and shifted SSP templates can be applied to describe the SDSS spectrum with 
emission lines being masked out by line width about 400km/s at zero intensity, similar as what we have 
recently done in \citet{zh19, zh22b, zh22c, zh23}. Left panels of Fig.~\ref{ssp} show examples on the SSP 
method determined the best descriptions and the corresponding line spectrum of a Type-2 AGN SDSS 0533-51994-0031 
(plate-mjd-fiberid). Here, line spectrum is calculated by the SDSS spectrum minus the SSP method determined 
stellar continuum.

	After subtractions of the stellar continuum, emission lines can be described by Gaussian functions. 
Here, the narrow emission lines of H$\alpha$, H$\beta$, [O~{\sc iii}]$\lambda4959, 5007$\AA~ doublet and 
[N~{\sc ii}]$\lambda6548,~6583$\AA~ doublet are mainly considered, in order to classify SDSS main galaxies 
by properties of the narrow emission line flux ratios in the BPT diagram of O3HB versus N2HA. Each Gaussian 
component is applied to describe each narrow emission line, besides the [O~{\sc iii}] doublet which are 
described by core plus extended broad Gaussian components probably related to shifted wings \citep{ref35, 
ref36, ref37}. Due to few effects of broad emission lines, there are no severe restrictions on the model 
parameters of each Gaussian component, besides the flux ratio of [O~{\sc iii}] ([N~{\sc ii}]) doublet fixed 
to the theoretical value of 3, and the emission line flux not smaller than zero. Then, through the 
Levenberg-Marquardt least-squares minimization method, the narrow emission lines can be measured. As examples, 
the right panels of Fig.~\ref{ssp} show the best-fitting results to the narrow emission lines in the line 
spectrum of the Type-2 AGN SDSS 0533-51994-0031 of which spectrum shown in the left panels. Here, one point 
should be noted. As shown in the example in Fig.~\ref{ssp}, almost all the collected objects have their 
[N~{\sc ii}] doublet and narrow Balmer lines to be described without considering extended components.

	Based on the measured reliable line parameters\footnote{Actually, there is a SDSS provided databases 
'GalSpecLine' (\url{https://skyserver.sdss.org/dr16/en/help/browser/browser.aspx?cmd=description+GalSpecLine+U}) 
including measured emission line parameters of all the main galaxies by MPA-JHU, as well described in 
\citet{br04, tr04, ka03a} and in \url{https://www.sdss.org/dr15/spectro/galaxy_mpajhu/}. The measured line 
parameters in this manuscript are well consistent with those in the databases 'GalSpecLine'. The same results 
can be found by applications of the line parameters in 'GalSpecLine'.}, the main galaxies with reliable 
measurements of the narrow emission lines are collected, based on the criteria that each measured line parameter 
is at least five times larger than its corresponding measured uncertainty and that the flux ratio (Balmer 
decrement) of the narrow H$\alpha$ to the narrow H$\beta$ is less than 6 to ignore effects of serious 
obscuration on the following results.

\begin{figure*}
\centering\includegraphics[width = 18cm,height=9cm]{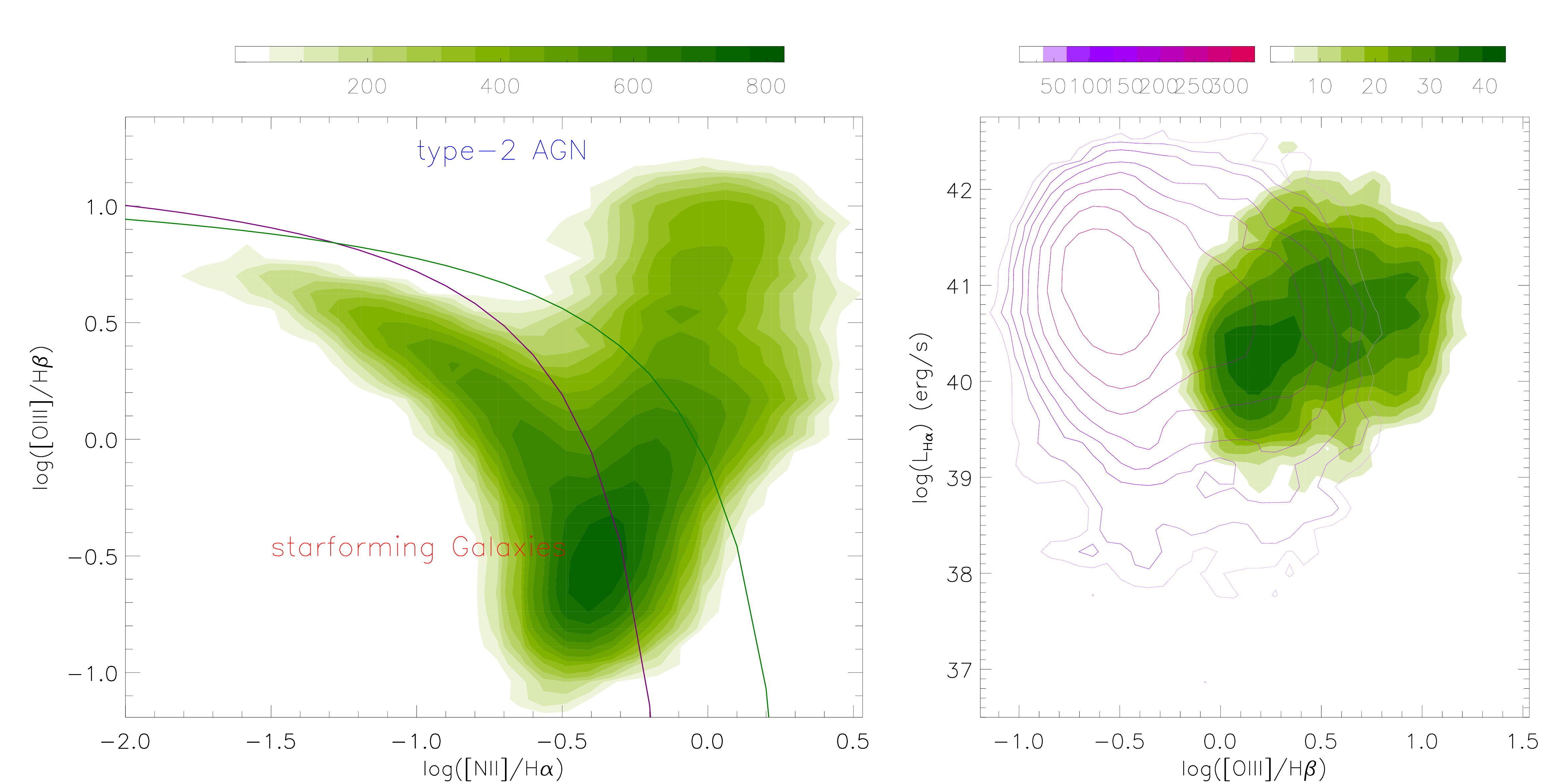}
\caption{Left panel shows the BPT diagram of O3HB versus N2HA for all the narrow emission-line main galaxies 
collected from the SDSS DR16, with the collected 4112 Type-2 AGN lying above the dividing line shown as solid 
dark green line and the collected 19351 starforming galaxies lying below the dividing line shown as solid purple 
line. Right panel shows the dependence of $L_{H\alpha}$ on the parameter O3HB. In the right panel, the contour 
filled with greenish colors represents the results in the Type-2 AGN, and the contour with contour levels in 
reddish lines represents the results in the starforming galaxies. In each panel, the color bar is shown in the 
top region to represent the corresponding number densities related to different colors.}
\label{bpt}
\end{figure*}

	Then, the well-known BPT diagram of O3HB versus N2HA is applied to classify the main galaxies into 
starforming galaxies and Type-2 AGN by the reported dividing lines in the literature \citep{ref20, ref21, ref22}.
\begin{equation}
\begin{split}
	&\log(O3HB)=0.61/(\log(N2HA)-0.05)+1.30 \ \ (green) \\
	&\log(O3HB)=0.61/(\log(N2HA)-0.47)+1.19 \ \ (purple)
\end{split}
\end{equation}.
As shown in the left panel of Fig.~\ref{bpt}, there are 19351 main galaxies classified as starforming galaxies 
in the BPT diagram lying below the dividing line shown as solid purple line, and 4112 main galaxies classified 
as Type-2 AGN in the BPT diagram lying above the dividing line shown as solid dark green line. Based on the 
measured line parameters, the median Balmer decrement is 4.10\ in the starforming galaxies and 4.18\ in the 
Type-2 AGN, indicating not different obscuration effects on the following results.

        Before ending the section, an additional point should be noted. Although the Type-2 AGN are collected
from the SDSS pipeline classified main galaxies (with no expected broad emission lines), it is necessary to 
check whether were there any collected Type-2 AGN with probably broad emission lines and probably AGN continuum 
emissions included in the SDSS spectra. After subtractions of the stellar continuum, emission lines around 
H$\alpha$ (rest wavelength from 6200 to 6800\AA) of all the collected 4112 Type-2 AGN have been re-measured by 
the narrow Gaussian functions applied to describe the narrow emission lines but by three additional broad 
Gaussian functions (second moment larger than 600km/s) applied to describe probably broad H$\alpha$. Based on 
the criterion that the measured parameters larger than 5 times of their corresponding uncertainties 
in one of the three broad Gaussian components in broad H$\alpha$, there are 285 Type-2 AGN with probably broad 
H$\alpha$. Therefore, there are 3836 (4121-285) Type-2 AGN in our final sample.

\section{Main results and Discussions}

\begin{figure*}
\centering\includegraphics[width = 18cm,height=8cm]{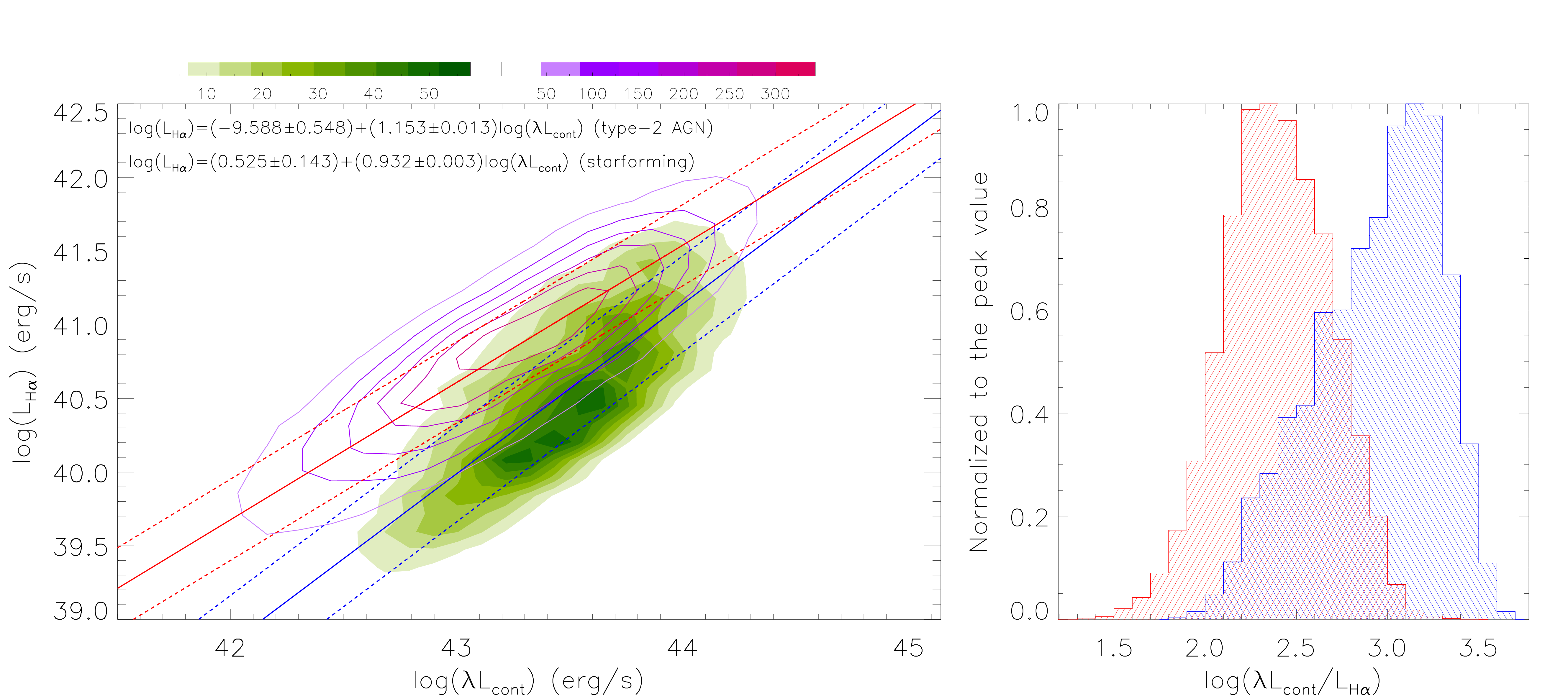}
\caption{Left panel shows the correlation between $L_{H\alpha}$ and $\lambda L_{cont}$ in the 19351 starforming 
galaxies (the contour with contour levels in reddish colors) and in the 3836 Type-2 AGN (the contour filled with 
greenish colors). Right panel shows the distributions of $\log(R_{CL})$ in the starforming galaxies (filled 
with red lines) and in the Type-2 AGN (filled with blue lines). In the left panel, solid and dashed blue lines, 
solid and dashed red lines show the best-fitting results (the corresponding formula marked in the top-left 
corner) and the corresponding 1RMS scatters to the correlations in the Type-2 AGN and in the starforming 
galaxies, respectively.
}
\label{hc}
\end{figure*}

	Based on the measured line parameters and the continuum emission features of the collected 19351 
starforming galaxies, there is a strong linear correlation between the narrow H$\alpha$ line luminosity 
($L_{H\alpha}$) and the continuum luminosity at 5100\AA~ ($\lambda L_{cont}$) with the Spearman rank correlation 
coefficient of 0.88 ($P_{null}<10^{-15}$). In this manuscript, although only objects are collected 
with flux ratio smaller than 6 of narrow H$\alpha$ to narrow H$\beta$, the $L_{H\alpha}$ and $\lambda L_{cont}$ 
have been reddening corrected for the objects with flux ratios of narrow H$\alpha$ to narrow H$\beta$ larger 
than 3.1, accepted 3.1 as the intrinsic flux ratio of narrow H$\alpha$ to narrow H$\beta$. After considering 
the uncertainties in both coordinates, the best fitting results shown in the left panel of Fig.~\ref{hc} are 
determined through the Least Trimmed Squares (LTS) robust technique 
\citep{ref23, ref24},
\begin{equation}
\log(\frac{L_{H\alpha}}{\rm erg\cdot s^{-1}})~=~(0.525\pm0.143)~+~
	(0.932\pm0.003)\log(\frac{\lambda L_{cont}}{\rm erg\cdot s^{-1}})
\end{equation}
The strong linear correlation in the starforming galaxies can be used as a standard candle to check effects of 
AGN feedback, by comparing the correlations of $L_{H\alpha}$ versus $\lambda L_{cont}$ between starforming 
galaxies and Type-2 AGN.

	Based on the measured parameters of the Type-2 AGN, there is also a strong linear correlation between 
$L_{H\alpha}$ and $\lambda L_{cont}$, with the Spearman rank correlation coefficient of 0.80 
($P_{null}<10^{-15}$), also shown in the left panel of Fig.~\ref{hc}. Here, the $L_{H\alpha}$ and 
$\lambda L_{cont}$ are the reddening corrected values in the Type-2 AGN. After considering the uncertainties 
in both coordinates, the best-fitting results in the Type-2 AGN are determined by the LTS technique,
\begin{equation}
\log(\frac{L_{H\alpha}}{\rm erg\cdot s^{-1}})~=~(-9.588\pm0.548)~+~
	(1.153\pm0.013)\log(\frac{\lambda L_{cont}}{\rm erg\cdot s^{-1}})
\end{equation}
It is clear that the linear correlations have much different intercepts but similar slopes between the Type-2 
AGN and the starforming galaxies.

	The expected $L_{H\alpha}$ are statistically smaller for given continuum luminosities in the Type-2 
AGN than in the starforming galaxies. Right panel of Fig.~\ref{hc} shows the distributions of luminosity ratio 
$R_{CL}$ of $\lambda L_{cont}$ to $L_{H\alpha}$. The median values of $\log(R_{CL})$ are about 2.39$\pm$0.29  
and 2.99$\pm$0.34\ in the starforming galaxies and in the Type-2 AGN, respectively. Uncertainties of the median 
values are the standard deviations of the $\log(R_{CL})$. The student's T-statistic technique can be applied 
to confirm that the starforming galaxies and the Type-2 AGN have different mean values of $\log(R_{CL})$ with 
significance level higher than 5$\sigma$. Therefore, the different correlations between $\lambda L_{cont}$ 
and $L_{H\alpha}$ can be confirmed in the starforming galaxies and in the Type-2 AGN.

\begin{figure*}
\centering\includegraphics[width = 18cm,height=8cm]{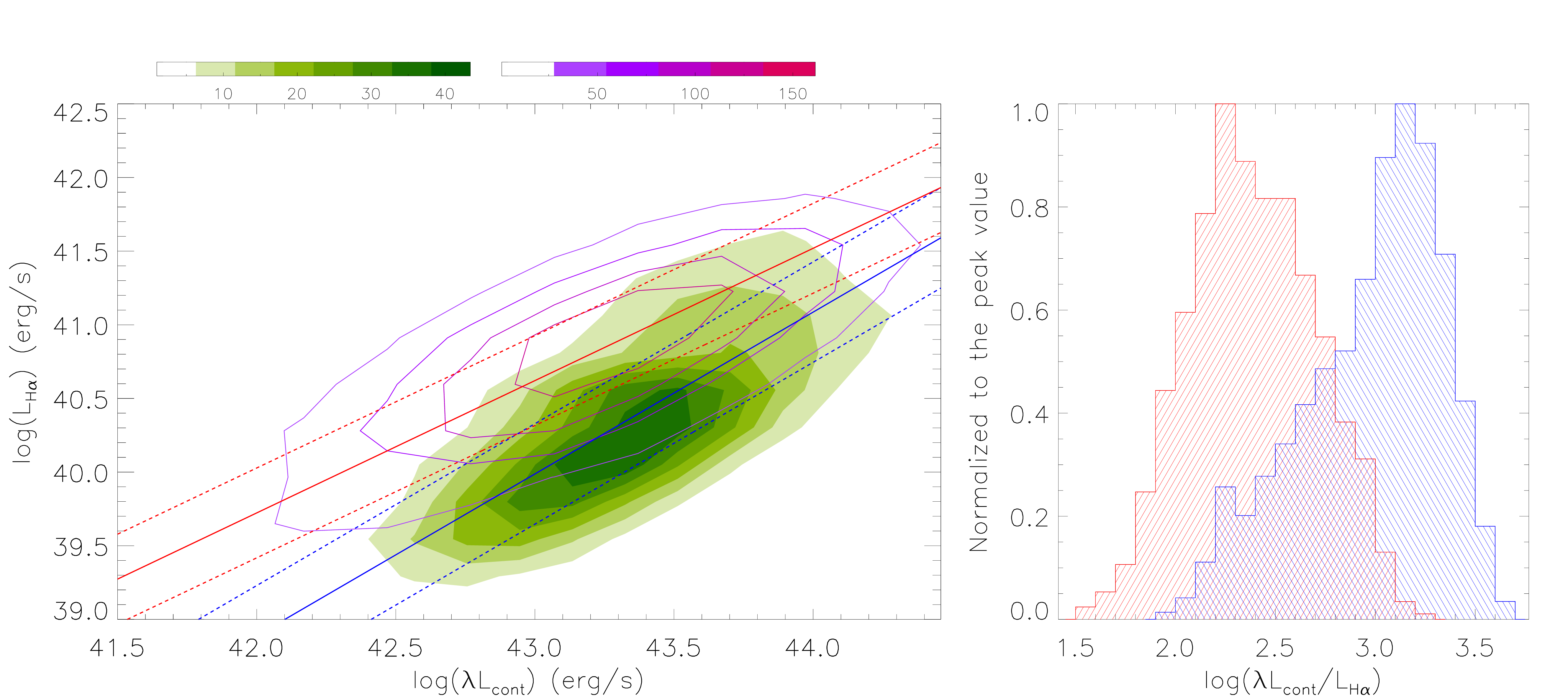}
\caption{Same as Fig.~\ref{hc}, but for the starforming galaxies with $S/N>30$ and the Type-2 AGN with $S/N>30$.}
\label{hc3}
\end{figure*}

	Before proceeding further, effects of signal-to-noise ratio ($S/N$) of SDSS spectra are simply discussed 
as follows on the shown results in Fig.~\ref{hc} for the starforming galaxies with $S/N>20$ and the Type-2 AGN 
with $S/N>20$. Among the starforming galaxies and the Type-2 AGN, the 2957 starforming galaxies with $S/N>30$ 
and the 1060 Type-2 AGN with $S/N>30$ are collected to re-check the correlation between $L_{H\alpha}$ versus 
$\lambda L_{cont}$, shown in Fig.~\ref{hc3}. The linear correlations can be confirmed with the Spearman rank 
correlation coefficients of about 0.84 ($P_{null}<10^{-15}$) and 0.79 ($P_{null}<10^{-15}$) for the 
starforming galaxies with $S/N>30$ and for the Type-2 AGN with $S/N>30$, respectively. And the best fitting 
results can be described as 
\begin{equation}
\begin{split}
	&\log(\frac{L_{H\alpha}}{\rm erg\cdot s^{-1}})~\propto~
	0.898\log(\frac{\lambda L_{cont}}{\rm erg\cdot s^{-1}}) \ \ \  (starforming)\\
	&\log(\frac{L_{H\alpha}}{\rm erg\cdot s^{-1}})~\propto~
	1.098\log(\frac{\lambda L_{cont}}{\rm erg\cdot s^{-1}}) \ \ \ (AGN)
\end{split}
\end{equation}
The median values and the corresponding standard deviations (as uncertainties) of $\log(R_{CL})$ are about 
2.37$\pm$0.31 and 3.04$\pm$0.35\ in the 2957 starforming galaxies with $S/N>30$ and in the 1060 Type-2 AGN 
with $S/N>30$, respectively. And the student's T-statistic technique can be applied to confirm the different 
mean values of $\log(R_{CL})$ with significance level higher than 5$\sigma$. The similar results as those 
shown in Fig.~\ref{hc} strongly indicate few effects of S/N on our final results. Therefore, there are no 
further discussions on effects of $S/N$.

	Considering the strong connections between starforming properties and narrow H$\alpha$ line luminosities, 
effects of AGN feedback on narrow H$\alpha$ line luminosity in Type-2 AGN can be expected. If there was positive 
AGN Feedback on starforming, statistically stronger narrow H$\alpha$ emissions could be expected in the Type-2 
AGN than in the starforming galaxies, otherwise, negative AGN feedback should lead to statistically weaker narrow 
H$\alpha$ emissions in the Type-2 AGN. Based on the results in Fig.~\ref{hc}, there are weaker narrow H$\alpha$ 
emissions (larger values of $R_{CL}$) in the Type-2 AGN than in the starforming galaxies. Therefore, the results 
in Fig.~\ref{hc} can be accepted as apparent clues to support negative AGN feedback in the local Type-2 AGN in 
SDSS, considering the continuum emissions with few contaminations of central AGN activity in host galaxies of 
Type-2 AGN. 

	In order to confirm the shown results in Fig.~\ref{hc} leading to negative AGN feedback, the following 
effects are mainly considered.

	If the lower line intensities of intrinsic narrow H$\alpha$ in Type-2 AGN was not due to the negative 
AGN Feedback but due to central AGN activity, it will be necessary to check whether stronger AGN activity can 
lead to lower line intensities of intrinsic narrow H$\alpha$. If lower line intensities of intrinsic narrow 
H$\alpha$ in Type-2 AGN were actually due to stronger AGN activity, statistically lower $L_{H\alpha}$ for given 
$\lambda L_{cont}$ could be expected in Type-2 AGN. However, in the collected Type-2 AGN, after checking the 
dependence of narrow H$\alpha$ luminosity on central AGN activity traced by the narrow emission line ratio of 
O3HB, one positive correlation can be found with the Spearman rank correlation coefficient of 0.31  
($P_{null}<10^{-15}$), shown in the right panel of Fig.~\ref{bpt}. Meanwhile, in the collected starforming 
galaxies, there is a weak negative dependence of $L_{H\alpha}$ on the parameter of O3HB, with the Spearman 
rank correlation coefficient of -0.22 ($P_{null}<10^{-15}$), also shown in the right panel of Fig.~\ref{bpt}. 
Therefore, the observed lower $L_{H\alpha}$ in Type-2 AGN is not due to stronger central AGN activity.

\begin{figure*}
\centering\includegraphics[width = 18cm,height=5.6cm]{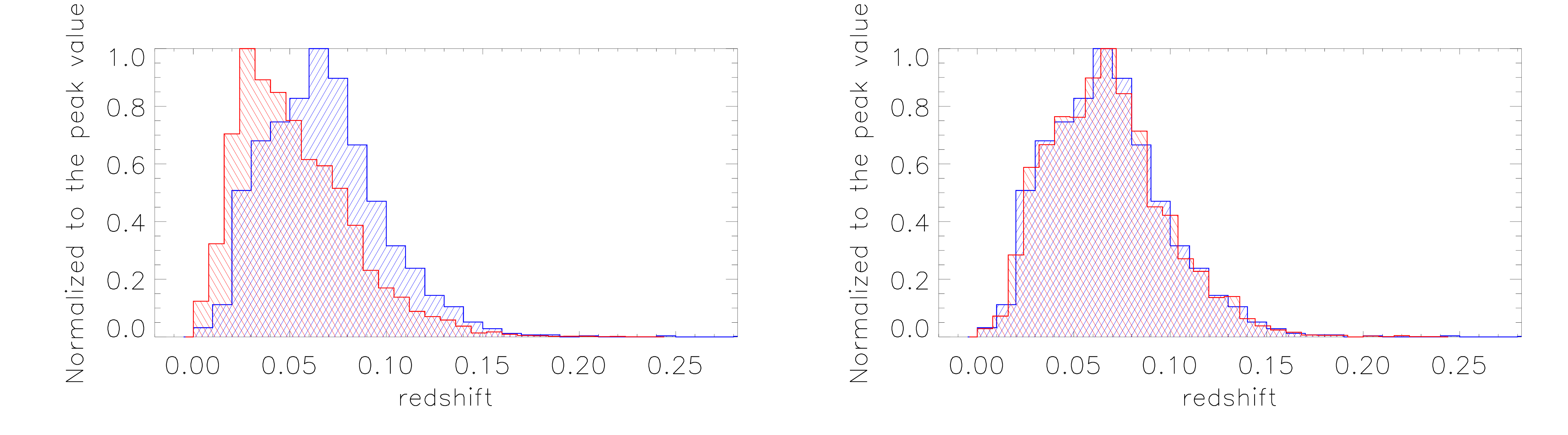}
\caption{Left panel shows the redshift distributions of the collected 19351 starforming galaxies (filled with 
red lines) and the collected 3836 Type-2 AGN (filled with blue lines). Right panel shows the redshift 
distributions of the randomly re-collected 7672 starforming galaxies (filled with red lines) in the subsample 
and the 3836 Type-2 AGN (filled with blue lines).
}
\label{zd}
\end{figure*}

	Furthermore, effects of the different redshift distributions are considered on the results shown in 
Fig.~\ref{hc} between the starforming galaxies and the Type-2 AGN. Left panel of Fig.~\ref{zd} shows the redshift 
distributions of the 19351 starforming galaxies with median $z$ about 0.047$\pm$0.029 and the 3836 Type-2 AGN with 
median $z$ about 0.065$\pm$0.030. Uncertainties of the median values are the standard deviations of $z$. And the 
student's T-statistic technique can be applied to confirm that the median values of $z$ are different with 
significance level higher than 5$\sigma$. Besides the different median redshifts, through the two-sided 
Kolmogorov-Smirnov statistic technique, the starforming galaxies and the Type-2 AGN have the same redshift 
distributions with significance level smaller than $10^{-15}$. Therefore, it is necessary to check effects of 
the different redshift distributions on the results shown in Fig.~\ref{hc}. In order to ignore the effects of 
the different redshift distributions, the simplest method is to check the results shown in Fig.~\ref{hc} but 
for two samples of starforming galaxies and Type-2 AGN which have the same redshift distributions.  

\begin{figure*}
\centering\includegraphics[width = 18cm,height=8cm]{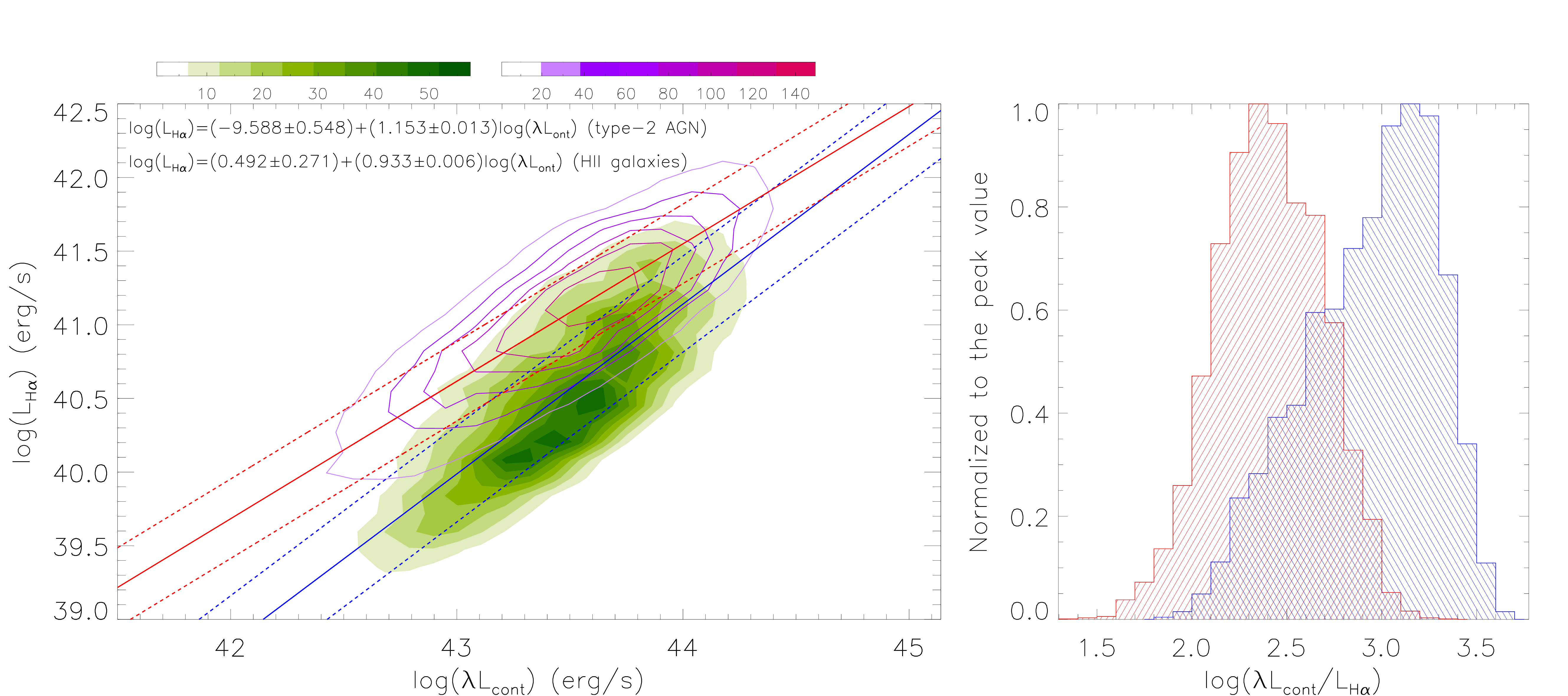}
\caption{Similar as Fig.~\ref{hc}, but the starforming galaxies are the re-collected 7672 starforming galaxies 
in the subsample.}
\label{h2c}
\end{figure*}

	Based on the redshift distributions of the starforming galaxies and the Type-2 AGN, 7672 starforming 
galaxies are easily and randomly collected into one subsample, which has the similar redshift distribution as 
that of the sample of the 3836 Type-2 AGN, as the results shown in the right panel of Fig.~\ref{zd}. Through 
the two-sided Kolmogorov-Smirnov statistic technique, the subsample of the 7672 starforming galaxies and the 
3836 Type-2 AGN have the same redshift distributions with significance level higher than 99.99\%. Then, the 
correlation between $L_{H\alpha}$ and $\lambda L_{cont}$ is shown in the left panel of Fig.~\ref{h2c} for the 
subsample of the 7672 starforming galaxies. The linear correlation can be confirmed with the Spearman rank 
correlation coefficient of about 0.88 ($P_{null}<10^{-15}$) for the re-collected 7672 starforming galaxies 
in the subsample. And the LTS technique determined best fitting results are shown in the left panel of 
Fig.~\ref{h2c} with the corresponding formula marked in the top region of the left panel of Fig.~\ref{h2c}. 
It is clear that the lower expected $L_{H\alpha}$ for given $\lambda L_{cont}$ can be re-confirmed in the 
Type-2 AGN than in the starforming galaxies in the subsample, as shown in the right panel of Fig.~\ref{h2c} 
with the median values and the corresponding standard deviations of $\log{R_{CL}}$ about 2.40$\pm$0.28 and 
2.99$\pm$0.34\ in the 7672 starforming galaxies in the subsample and in the 3836 Type-2 AGN, respectively. 
And the student's T-statistic technique can be applied to confirm the different median values of $\log{R_{CL}}$ 
with significance level higher than 5$\sigma$. Therefore, the very different correlations between $L_{H\alpha}$ 
and $\lambda L_{cont}$ are intrinsic and reliable between the starforming galaxies and the Type-2 AGN, after 
considering effects of different $z$ distributions of the starforming galaxies and the Type-2 AGN.

\begin{figure*}
\centering\includegraphics[width = 18cm,height=5cm]{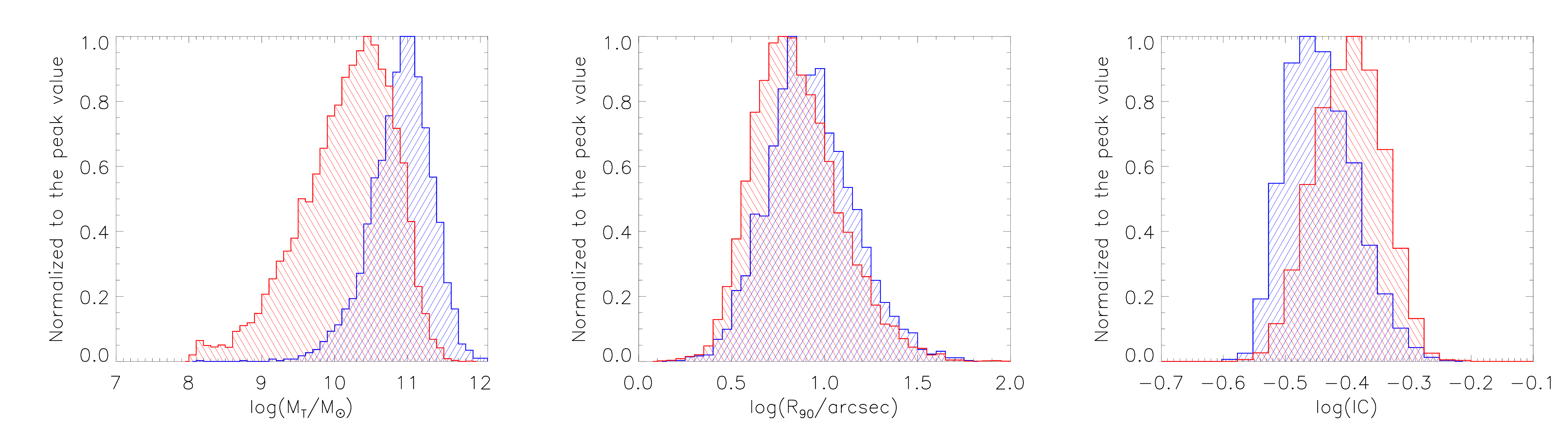}
\caption{Left panel shows the $log(M_T)$ distributions of the collected 18995 starforming galaxies (filled 
with red lines) and the collected 3827 Type-2 AGN (filled with blue lines). Middle and right panel show 
the distributions of the $R_{90}$ and $IC$ of the 18131 starforming galaxies and the 3652 Type-2 AGN with 
reliable measurements of parameters of petroR90\_r and petroR50\_r.}
\label{md}
\end{figure*}

\begin{figure*}
\centering\includegraphics[width = 18cm,height=12cm]{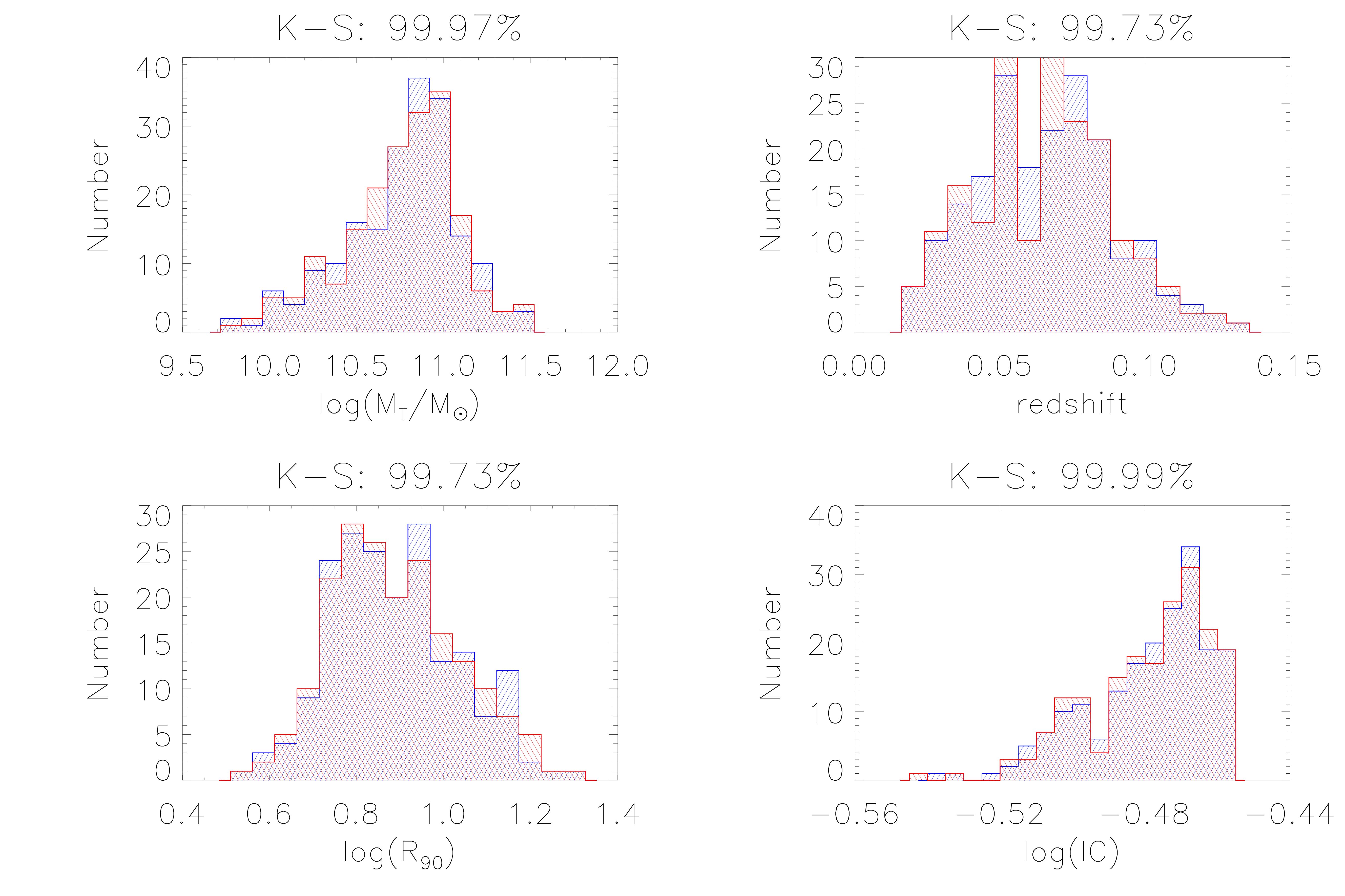}
\caption{Distributions of the $M_T$ (top left panel), $z$ (top right panel), $R_{90}$ (bottom left panel) and 
$IC$ (bottom right panel) of the 191 starforming galaxies and the 191 Type-2 AGN in the subsamples with the 
same distributions of $M_T$, $z$, $R_{90}$ and $IC$ ($IC~<~0.35$ and devab\_r larger than 0.8). In each panel, 
histogram filled with red lines shows the results for the starforming galaxies, histogram filled with 
blue lines shows the results for the Type-2 AGN.}
\label{mor}
\end{figure*}

\begin{figure*}
\centering\includegraphics[width = 18cm,height=8cm]{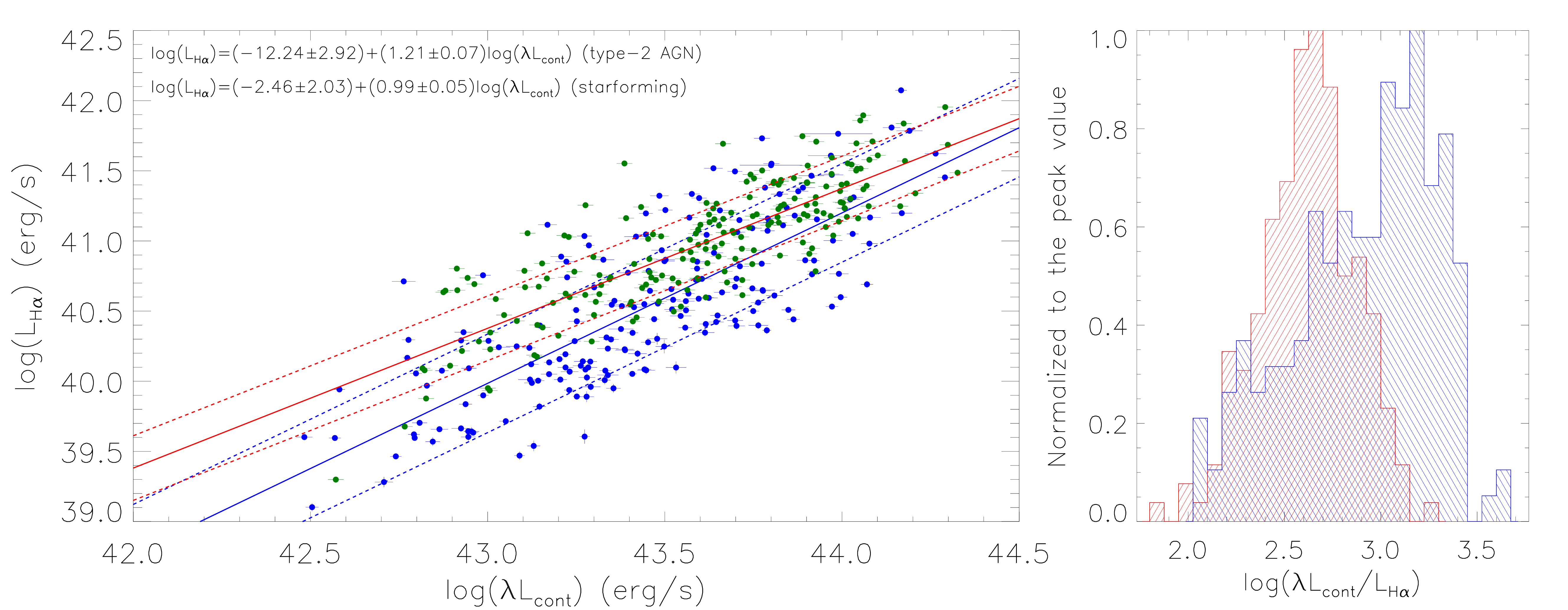}
\caption{Similar as Fig.~\ref{hc} and as Fig.~\ref{h2c}, but for the re-collected 191 starforming galaxies 
and the re-collected 191 Type-2 AGN in the subsamples which have the same distributions of $M_T$, $z$, $_{90}$ 
and $IC$ ($IC<0.35$ and devab\_r larger than 0.8). In the left panel, sold circles plus error bars in blue 
show the results for the 191 Type-2 AGN, and circles plus error bars in dark green show the results for the 
191 starforming galaxies.}
\label{mt}
\end{figure*}

\begin{figure*}
\centering\includegraphics[width = 18cm,height=6cm]{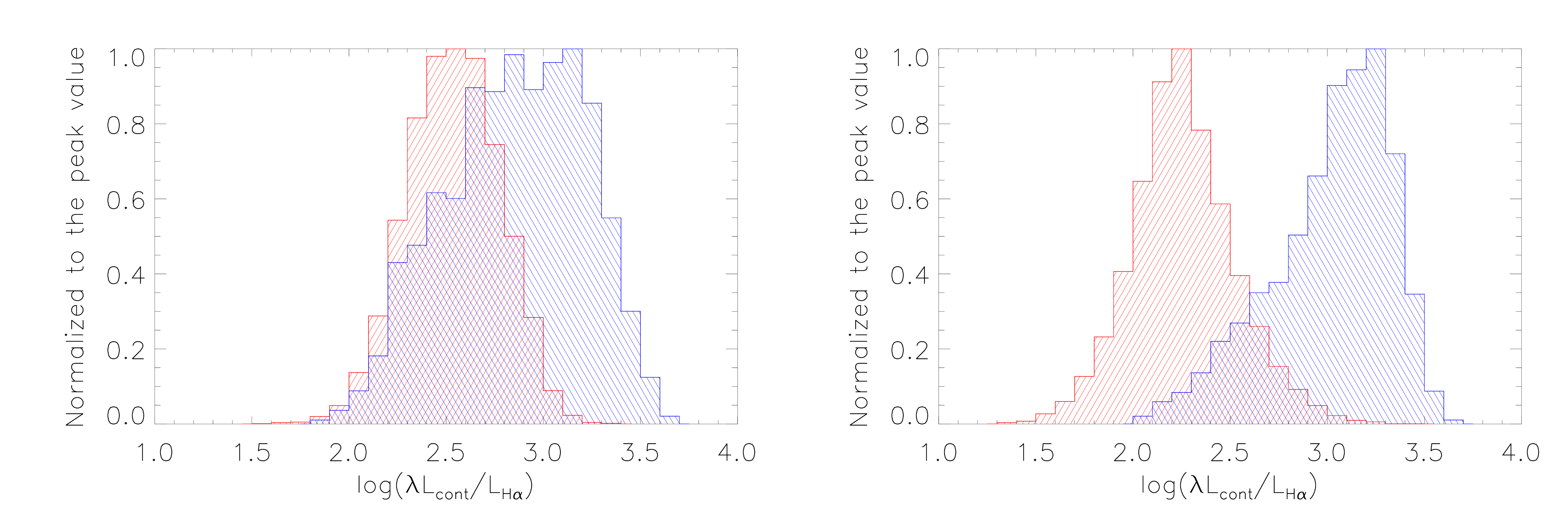}
\caption{Left panel shows the distributions of $\log{R_{CL}}$ of the starforming galaxies in the SF\_H sample 
and the Type-2 AGN in the AGN\_L sample. Right panel shows the distributions of $\log{R_{CL}}$ of the 
starforming galaxies in the SF\_L sample and the Type-2 AGN in the AGN\_H sample. In each panel, histogram 
filled with dark green lines shows the results for the starforming galaxies, histogram filled with blue 
lines shows the results for the Type-2 AGN.}
\label{HL}
\end{figure*}

	Furthermore, combining with aperture effects, effects of the different distributions of total stellar 
mass $M_{T}$ and different host galaxy morphologies are considered on the results shown in Fig.~\ref{hc} 
between the starforming galaxies and the Type-2 AGN. Left panel of Fig.~\ref{md} shows the $M_{T}$ 
distributions of the 18995 starforming galaxies (356 starforming galaxies not included, due to their total 
stellar masses smaller than five times of their corresponding uncertainties) with median 
$\log(M_{T}/{\rm M_\odot})$ about 10.26$\pm$0.63 and of the 3827 Type-2 AGN (36 Type-2 AGN not included, 
due to their total stellar masses smaller than five times of their corresponding uncertainties) with median 
$\log(M_{T}/{\rm M_\odot})$ about 10.94$\pm$0.41. Uncertainties of the median values are the standard 
deviations of $\log(M_{T}/{\rm M_\odot})$. And the student's T-statistic technique can be applied to confirm 
that the median values of $\log(M_{T})$ are different with significance level higher than 5$\sigma$. And, 
through the two-sided Kolmogorov-Smirnov statistic technique, the starforming galaxies and the Type-2 AGN 
have the same redshift distributions with significance level smaller than $10^{-15}$. Therefore, it is 
necessary to check effects of the different $M_{T}$ distributions on the results shown in Fig.~\ref{hc}. 

	Meanwhile, middle panel of Fig.~\ref{md} shows the $R_{90}$ (as the radii containing 90\% of the 
Petrosian flux in SDSS r band\footnote{The same results can be found, if the Petrosian radius applied in 
the other bands.}) distributions of the 18131 starforming galaxies (1219 starforming galaxies not included, 
due to their petroR90\_r smaller than five times of their corresponding uncertainties) with median 
$\log(R_{90}/arcsec)$ about 0.83$\pm$0.23 and of the 3652 Type-2 AGN (184 Type-2 AGN not included, due to 
their petroR90\_r smaller than five times of their corresponding uncertainties) with median 
$\log(R_{90}/arcsec)$ about 0.91$\pm$0.22. Uncertainties of the median values are the standard deviations 
of $\log(R_{90}/arcsec)$. And the student's T-statistic technique can be applied to confirm that the median 
values of $\log(R_{90})$ are different with significance level higher than 5$\sigma$. And, through the 
two-sided Kolmogorov-Smirnov statistic technique, the starforming galaxies and the Type-2 AGN have the same 
$R_{90}$ distributions with significance level smaller than $10^{-15}$. Therefore, it is necessary to check 
effects of the different $R_{90}$ distributions on the results shown in Fig.~\ref{hc}.

	Meanwhile, right panel of Fig.~\ref{md} shows the distributions of the inverse concentration parameter 
\citep{ref29, ref30} $IC=R_{50}/R_{90}$ (where $R_{50}$ and $R_{90}$ as the radii containing 50\% and 90\% of 
the Petrosian flux in SDSS r band\footnote{The same results can be found, if the Petrosian radius applied in 
the other bands.}) of the 18131 starforming galaxies (with reliable measurements of petroR90\_r and petroR50\_r) 
and the 3652 Type-2 AGN (with reliable measurements of petroR90\_r and petroR50\_r). The median values of 
$\log(IC)$ are about -0.39$\pm$0.05 and -0.45$\pm$0.06 for the starforming galaxies and the Type-2 AGN, 
respectively. And the student's T-statistic technique can be applied to confirm that the median values of 
$\log(IC)$ are different with significance level higher than 5$\sigma$. And, through the two-sided 
Kolmogorov-Smirnov statistic technique, the starforming galaxies and the Type-2 AGN have the same $IC$ 
distributions with significance level smaller than $10^{-15}$. Therefore, it is necessary to check effects 
of the different $IC$ distributions on the results shown in Fig.~\ref{hc}.

	In order to check effects of different distributions of $z$, $M_T$, $R_{90}$ and $IC$, a simple method 
can be considered as follows by re-collected starforming galaxies and Type-2 AGN into new subsamples which have 
the same distributions of $z$, $M_T$, $R_{90}$ and $IC$. The same distributions of $z$ and $R_{90}$ can be 
applied to ignore aperture effects on the results for the objects in the subsamples. The same distribution of 
$M_T$ can be applied to ignore effects of different total stellar mass on the results for the objects in the 
subsamples. And the same distributions of $IC<0.35$ (combining with devab\_r larger than 0.8) can be applied to 
simply accepted that the host galaxies of the re-collected objects are elliptical galaxies as discussed 
in \citet{ref29, ref30} and in \url{https://www.sdss4.org/dr16/algorithms/classify/}, in order to ignore 
effects of inclinations of disk galaxies and/or to ignore probable effects of different morphologies. Based 
on the $z$ distributions (shown in the left panel of Fig.~\ref{zd}) and the distributions of $M_T$, $R_{90}$ 
and $IC$ shown in Fig.~\ref{md} of the starforming galaxies and the Type-2 AGN, a subsample of 191 starforming 
galaxies and a subsample of 191 Type-2 AGN can be collected, with the two subsamples having the same 
distributions of $z$, $M_T$, $R_{90}$ and $IC$, which are shown in Fig.~\ref{mor}. Through the two-sided 
Kolmogorov-Smirnov statistic technique, the subsample of the 191 starforming galaxies and the subsample of 
the 191 Type-2 AGN have the same distributions of $z$, $M_T$, $R_{90}$ and $IC$, with significance levels 
higher than 99.5\%. 

	Then, the correlations between $L_{H\alpha}$ and $\lambda L_{cont}$ are shown in the left panel of 
Fig.~\ref{mt} for the 191 starforming galaxies and the 191 Type-2 AGN in the subsamples which have the same 
distributions of $z$, $M_T$, $R_{90}$ and $IC$. And the mean ratio of emission line flux to its corresponding 
uncertainty is about 36\ in narrow H$\alpha$, the mean ratio of continuum emission intensity to its 
corresponding uncertainty is about 21. The linear correlations can be confirmed with the Spearman rank 
correlation coefficients of about 0.83 ($P_{null}<10^{-15}$) for the re-collected 191 starforming galaxies 
in the subsample and of about 0.78 ($P_{null}<10^{-15}$) for the re-collected 191 Type-2 AGN in the subsample. 
And the LTS technique determined best fitting results are shown in the left panel of Fig.~\ref{mt} with the 
corresponding formula marked in the top region of the left panel of Fig.~\ref{mt}. It is clear that the lower 
expected $L_{H\alpha}$ for given $\lambda L_{cont}$ can be re-confirmed in the Type-2 AGN than in the starforming 
galaxies in the subsamples, as shown in the right panel of Fig.~\ref{mt} with the median values and the 
corresponding standard deviations of $\log{R_{CL}}$ about 2.61$\pm$0.24 and 2.98$\pm$0.36\ in the 191  
starforming galaxies and the 191 Type-2 AGN in the subsamples, respectively. And the student's T-statistic 
technique can be applied to confirm the different median values of $\log{R_{CL}}$ with significance levels 
higher than 5$\sigma$. Therefore, even after considering effects of different distributions of $z$, $M_T$, 
$R_{90}$ and $IC$, the very different correlations between $L_{H\alpha}$ and $\lambda L_{cont}$ are intrinsic 
and reliable between the starforming galaxies and the Type-2 AGN. 

	Moreover, due to unconfirmed positive or negative AGN feedback on starforming histories, the following 
extreme comparisons can be checked between starforming galaxies with lower (higher) total stellar masses and 
Type-2 AGN with higher (lower) total stellar masses. Among the 18995 starforming galaxies with median 
$\log(M_{T}/{\rm M_\odot})$ about 10.26, there are half of the starforming galaxies in SF\_H sample with 
total stellar masses $\log(M_{T}/{\rm M_\odot})$ larger than the median value 10.26, and the other half of 
the starforming galaxies in SF\_L sample with total stellar masses $\log(M_{T}/{\rm M_\odot})$ smaller than 
the median value 10.26. Meanwhile, among the 3836 Type-2 AGN with median $\log(M_{T}/{\rm M_\odot})$ about 
10.94, there are half of the Type-2 AGN in AGN\_H sample with total stellar masses $\log(M_{T}/{\rm M_\odot})$ 
larger than the median value 10.94, and the other half of the Type-2 AGN in AGN\_L sample with total stellar 
masses $\log(M_{T}/{\rm M_\odot})$ smaller than the median value 10.94. Then, left panel of Fig.~\ref{HL} 
shows the distributions of $\log{R_{CL}}$ of the 9478 starforming galaxies in the SF\_H sample with median 
$\log{R_{CL}}$ about 2.53$\pm$0.24 and the 1918 Type-2 AGN in the AGN\_L sample with median $\log{R_{CL}}$ 
about 2.88$\pm$0.35, right panel of Fig.~\ref{HL} shows the distributions of $\log{R_{CL}}$ of the 9477 
starforming galaxies in the SF\_L sample with median $\log{R_{CL}}$ about 2.25$\pm$0.26 and the 1918 Type-2 
AGN in the AGN\_L sample with median $\log{R_{CL}}$ about 3.07$\pm$0.30. And the student's T-statistic 
technique can be applied to confirm the different median values of $\log{R_{CL}}$ shown in Fig.~\ref{HL} with 
significance levels higher than 5$\sigma$. Therefore, the very different correlations between $L_{H\alpha}$ 
and $\lambda L_{cont}$ are intrinsic and reliable between the starforming galaxies and the Type-2 AGN, 
even after considering very different $M_T$ distributions of the starforming galaxies and the Type-2 AGN.

\begin{figure}
\centering\includegraphics[width = 12cm,height=8cm]{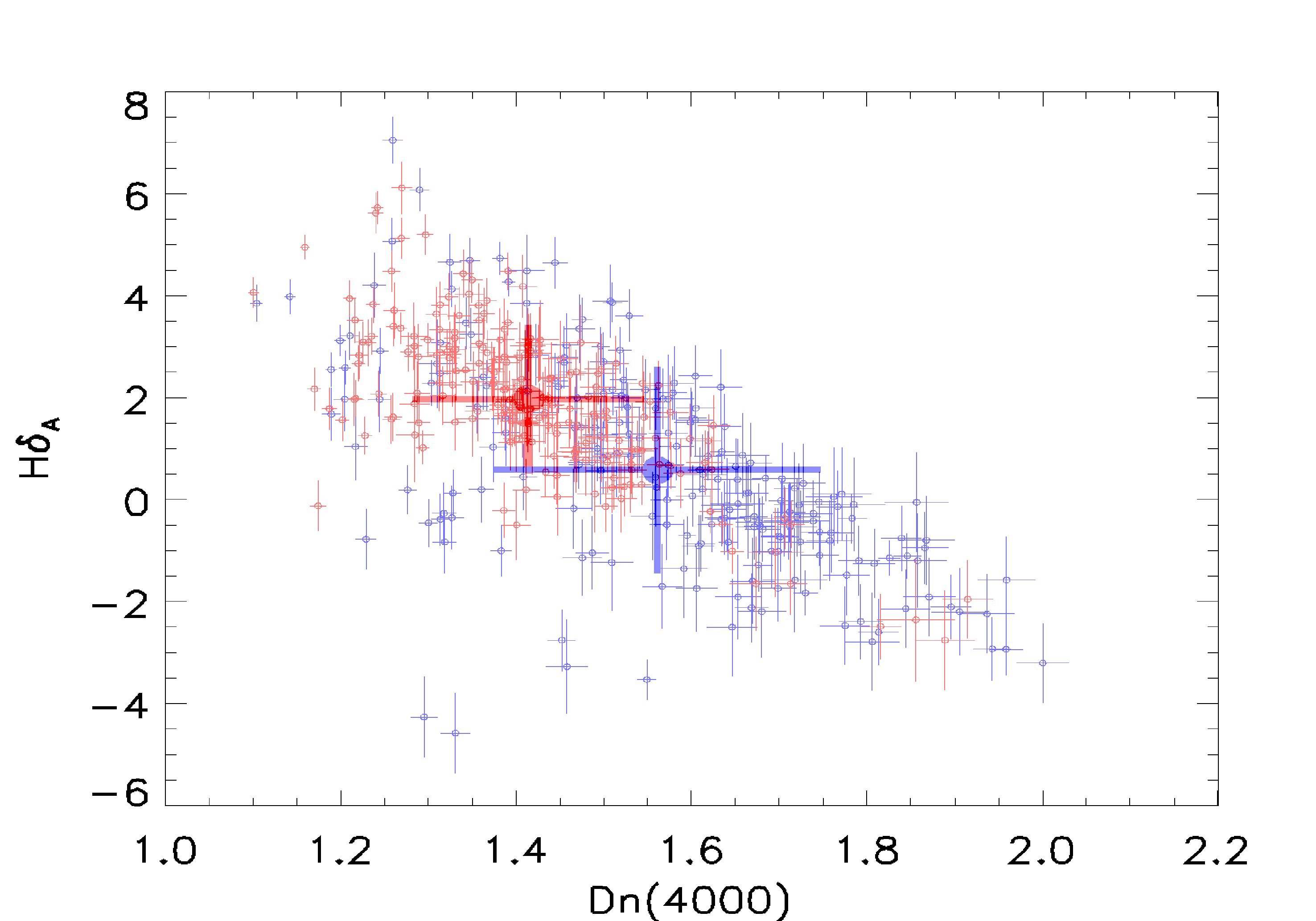}
\caption{On the properties of Dn(4000) and H$\delta_A$ in the 191 starforming galaxies (open circles plus 
error bars in red) and the 191 Type-2 AGN (open circles plus error bars in blue) in the subsamples 
with the same distributions of $z$, $M_T$, $R_{90}$ and $IC$ ($IC<0.35$ and devab\_r larger than 0.8). 
Solid circles plue error bars in red and in blue show the mean positions and the corresponding uncertainties 
(the standard deviation) of the starforming galaxies and the Type-2 AGN, respectively.}
\label{d4}
\end{figure}

	Moreover, the parameters of Dn(4000) and H$\delta_A$ \citep{ref27, ref28} are compared between the 
206 starforming  galaxies and the 206 Type-2 AGN in the subsamples with the same distributions of $z$, $M_T$, 
$R_{90}$ and $IC$ ($IC<0.35$ and devab\_r larger than 0.8), because the two parameters can provide intrinsic 
information on stellar ages \citep{ref21}. Here, the parameters Dn(4000) and H$\delta_A$ are measured through 
the spectroscopic features similar as what have been done in \citet{ref21}. The results on Dn(4000) and 
H$\delta_A$ are shown in Fig.~\ref{d4}. Median values and the corresponding standard deviations of Dn(4000) 
and H$\delta_A$ are 1.57$\pm$0.17 and 0.24$\pm$2.02\ in the 206 Type-2 AGN in the subsample, and 1.41$\pm$0.13 
and 1.99$\pm$1.45\ in the 206 starforming galaxies in the subsample, indicating older stellar populations 
in the host galaxies in the Type-2 AGN. And the student's T-statistic technique can be applied to confirm 
that the mean values of Dn(4000) (H$\delta_A$) are different with significance levels higher than 5$\sigma$, 
between the starforming galaxies and the Type-2 AGN in the subsamples with the same distributions of $z$, 
$M_T$, $R_{90}$ and $IC$ ($IC<0.35$ and devab\_r larger than 0.8). If there were positive AGN feedback on 
starforming in the host galaxies of Type-2 AGN, younger stellar ages should be clearly expected. Therefore, 
the results shown in Fig.~\ref{d4} can be accepted as indirect evidence to support the globally negative AGN 
feedback on starforming.

\begin{figure*}
\centering\includegraphics[width = 18cm,height=8cm]{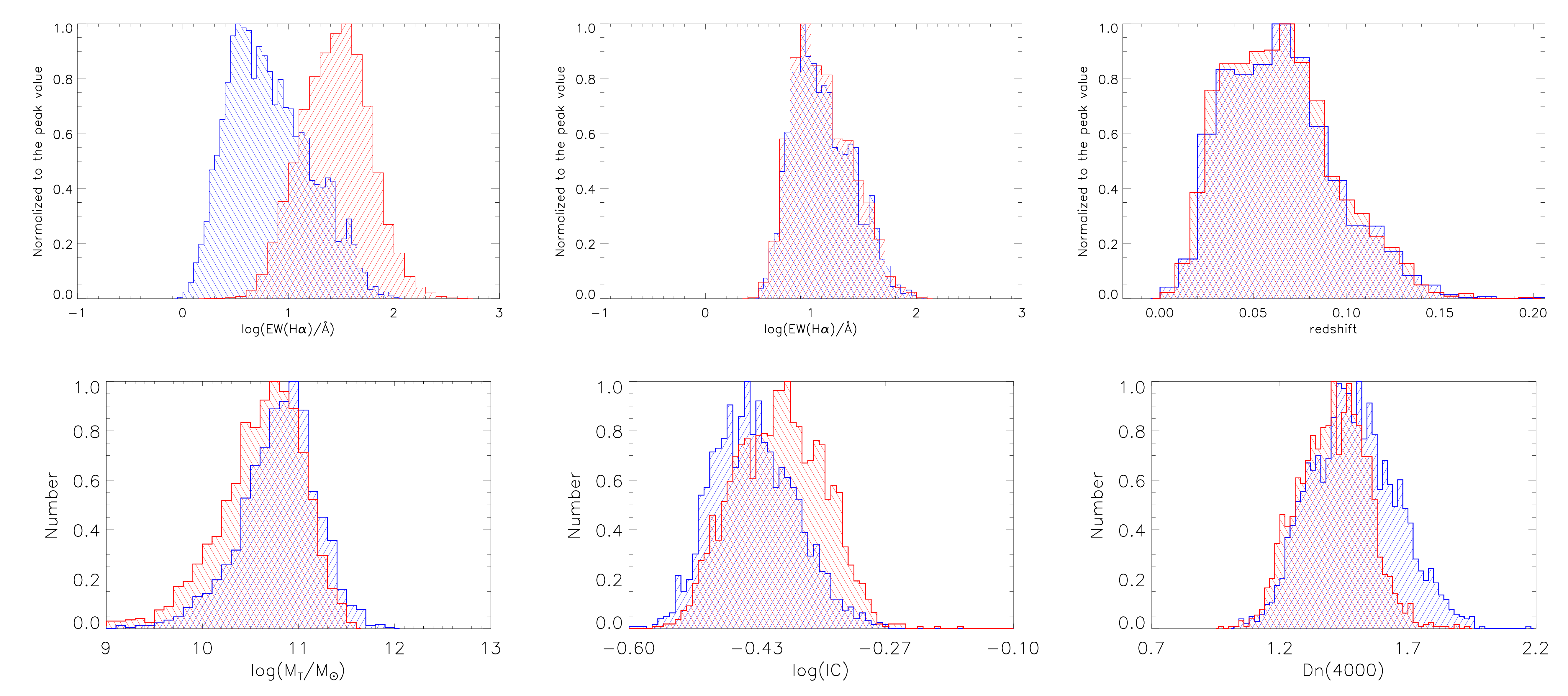}
\caption{Top left panel shows the Distributions of EW(H$\alpha$) of the starforming galaxies 
and the Type-2 AGN. Top middle panel and top right panel show the distributions of redshift and EW(H$\alpha$) 
of the starforming galaxies and the Type-2 AGN in the subsamples which have the same distributions of 
$z$ and EW(H$\alpha$). Bottom panels show the distributions of $M_T$, $IC$ and Dn(4000) of the starforming 
galaxies and the Type-2 AGN in the subsamples which have the same distributions of $z$ and EW(H$\alpha$). 
In each panel, histogram filled with red lines shoes the results for the starforming galaxies, and histogram 
filled with blue lines are for the Type-2 AGN.}
\label{ew}
\end{figure*}

\begin{figure*}
\centering\includegraphics[width = 18cm,height=4cm]{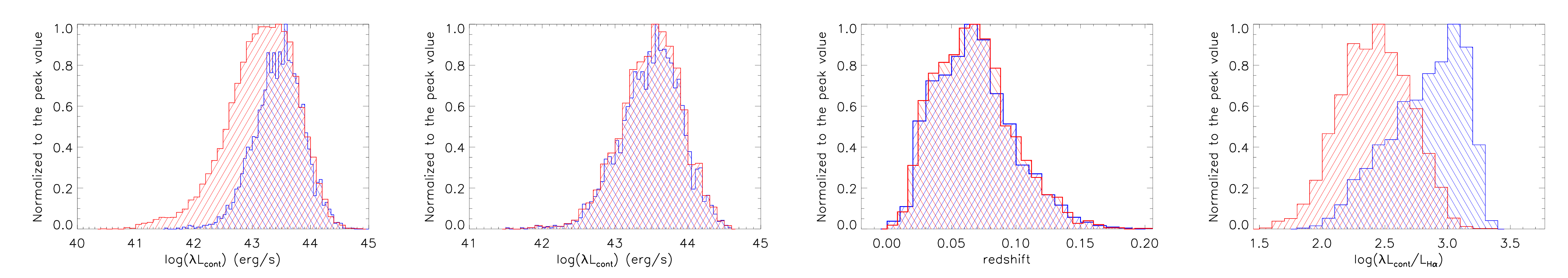}
\caption{Left panel shows the Distributions of $\lambda L_{cont}$ of the starforming galaxies 
and the Type-2 AGN. The second panel and the third panel show the distributions of redshift and 
$\lambda L_{cont}$ of the starforming galaxies and the Type-2 AGN in the subsamples which have the same 
distributions of $z$ and $\lambda L_{cont}$. Right panel shows the $\log(R_{CL})$ distributions of the 
starforming galaxies and the Type-2 AGN in the subsamples which have the same distributions of $z$ and 
$\lambda L_{cont}$. In each panel, histogram filled with red lines shoes the results for the starforming 
galaxies, and histogram filled with blue lines are for the Type-2 AGN.} 
\label{e2w}
\end{figure*}

	Furthermore, as discussed in \citet{cs11} that the Type-2 AGN and the starforming galaxies can be 
separated in the space of EW(H$\alpha$) (equivalent width of narrow H$\alpha$) and N2HA, therefore it is 
necessary to check whether the different $R_{CL}$ are only due to very different distributions of EW(H$\alpha$) 
between the starforming galaxies and the Type-2 AGN. Top left panel of Fig.~\ref{ew} shows the 
$\log(EW(H\alpha))$ distributions, with mean value of about 1.45$\pm$0.32 for the starforming galaxies 
and of about 0.83$\pm$0.38 for the Type-2 AGN. Then, based on the EW(H$\alpha$) and redshift distributions, 
one subsample including 2013 starforming galaxies and one subsample including 2013 Type-2 AGN are created, 
with the same distributions of $z$ and EW(H$\alpha$) as shown in the top middle panel and the top right panel 
of Fig.~\ref{ew}. Through the two-sided Kolmogorov-Smirnov statistic technique, the two subsamples have the 
same distributions of $z$ and EW(H$\alpha$) with significance levels higher than 99.99\%. Here, 
due to the same EW(H$\alpha$) distributions for the 2013 Type-2 AGN and the 2013 HII galaxies in the 
subsamples, the similar dependence dependence of $L_{H\alpha}$ on continuum luminosity can be expected,
because both the EW(H$\alpha$) and the linear dependence of $L_{H\alpha}$ on continuum luminosity have the 
same physical meanings. However, if there were no effects of AGN feedback on host galaxy properties of Type-2 
AGN, similar host galaxy properties could be expected between the 2013 Type-2 AGN and the 2013 HII galaxies 
in the subsamples. Nevertheless, through the shown distributions of $M_{T}$, $IC$ and Dn(4000) of the objects 
in the subsamples in bottom panels of Fig.~\ref{ew}, there are very different host galaxy properties between 
the starforming galaxies and the Type-2 AGN in the subsamples. The mean values of $\log(M_{T}/{\rm M_\odot})$, 
$\log(IC)$ and Dn(4000) are about 10.78$\pm$0.41, -0.44$\pm$0.05 and 1.49$\pm$0.16 for the 2013 Type-2 AGN 
in the subsample, and about 10.56$\pm$0.54, -0.40$\pm$0.06 and 1.41$\pm$0.13 for the 2013 starforming galaxies 
in the subsample. And the student's T-statistic technique can be applied to confirm that the mean values of
host galaxy properties are different enough with significance levels higher than 5$\sigma$. And through the 
two-sided Kolmogorov-Smirnov statistic technique, the significance levels smaller than $10^{-15}$ for the 
objects in the subsamples having the same distributions of $M_{T}$, $IC$ and Dn(4000). In one word, 
besides the different EW(H$\alpha$) distributions, AGN feedback could play the key role leading to the 
different $R_{CL}$. 

	Meanwhile, as shown above on the steeper dependence of $L_{H\alpha}$ and $\lambda L_{cont}$ in the 
Type-2 AGN, it is necessary to check whether more luminous Type-2 AGN included in our final sample can affect 
the statistical results on $R_{CL}$. Left panel of Fig.~\ref{e2w} shows the $\log(\lambda L_{cont}/(\rm erg/s))$ 
distributions with the mean value of about 43.15$\pm$0.59 for the starforming galaxies and of about 43.45$\pm$042 
for the Type-2 AGN. Then, based on the $\log(\lambda L_{cont})$ and redshift distributions, one subsample 
including 2967 starforming galaxies and one subsample including 2967 Type-2 AGN are created, with the same 
distributions of $z$ and $\log(\lambda L_{cont})$ as shown in the second panel and the third panel of 
Fig.~\ref{e2w}. Through the two-sided Kolmogorov-Smirnov statistic technique, the two subsamples have the 
same distributions of $z$ and $\log(\lambda L_{cont})$ with significance levels higher than 99.99\%. Then, 
right panel of Fig.~\ref{e2w} shows the distributions of $\log(R_{CL})$, with the mean value of about 
2.42$\pm$0.28 for the starforming galaxies and of about 2.83$\pm$0.29 for the Type-2 AGN in the subsamples. 
And the student's T-statistic technique can be applied to confirm that the mean values of $\log(R_{CL})$ 
are different enough with significance levels higher than 5$\sigma$. In one word, more luminous Type-2 AGN 
cannot be applied to explain the apparent different $\log(R_{CL})$ between the starforming galaxies and 
the Type-2 AGN.

\begin{figure*}
\centering\includegraphics[width = 18cm,height=9cm]{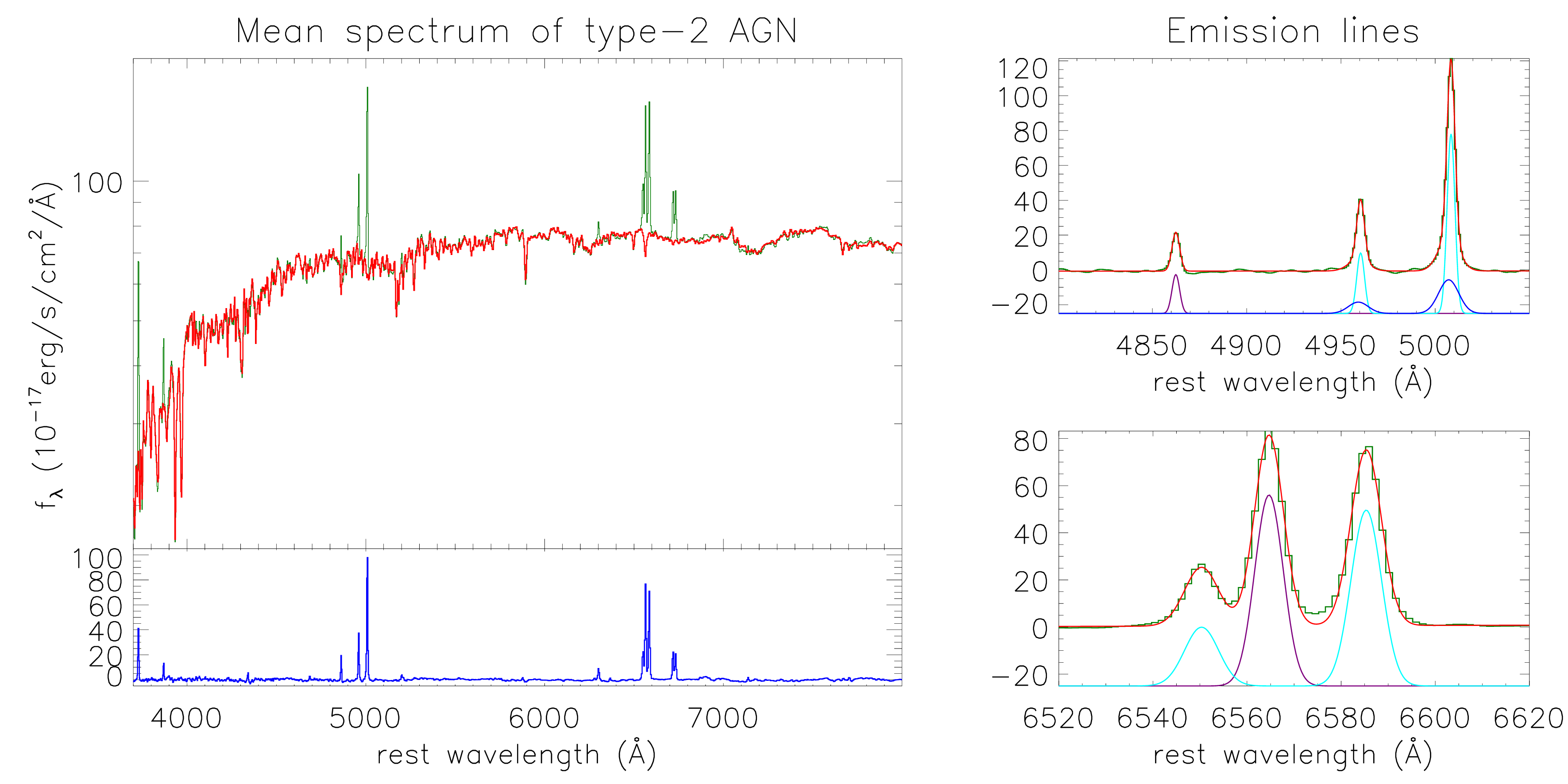}
\caption{The mean spectrum of the Type-2 AGN. The results are the same as those shown in Fig.~\ref{ssp}, but 
for the flux-weighted mean spectrum of the 4112 Type-2 AGN.}
\label{msp}
\end{figure*}

	Certainly, in this manuscript, it has been accepted that there are no contribution of central AGN 
emissions to continuum emissions of host galaxies of Type-2 AGN. If considering the AGN contributions to 
continuum emissions leading to the different correlations between $L_{H\alpha}$ and $\lambda L_{cont}$ in 
the starforming galaxies and in the Type-2 AGN, about 58\% of continuum emissions from central AGN activity 
could be expected in Type-2 AGN. So larger contributions should indicate stronger power law continuum emission 
components in SDSS spectra of Type-2 AGN. However, after checking SDSS spectra of the Type-2 AGN, there are 
no broad emission lines nor power law continuum components detected in SDSS spectra of Type-2 AGN. The 
flux-weighted mean spectrum of Type-2 AGN is shown in Fig.~\ref{msp}, which can be described by pure stellar 
templates without considerations of power law components, and have no apparent broad emission lines, to 
simply support that only small number of Type-2 AGN have broad emission lines and AGN continuum emissions 
included in the SDSS spectra. Furthermore, if there were strong contributions of central AGN activity to 
continuum emissions in host galaxies of Type-2 AGN, the calculated H$\delta_A$ could be about 2 times 
different from the shown results in left panel of Fig.~\ref{d4} in the Type-2 AGN, leading to unexpected quite 
different correlations between Dn(4000) and H$\delta_A$ in starforming galaxies and in host galaxies of Type-2 AGN.

\section{Summary and Conclusions}

The final summary and conclusions are as follows.
\begin{itemize}
\item Among all the low-redshift narrow emission line main galaxies in SDSS DR16 with $S/N>20$, the spectroscopic 
	narrow emission lines of H$\beta$, [O~{\sc iii}], H$\alpha$ and [N~{\sc ii}] are well measured, after 
	subtractions of host galaxy contributions determined by the SSP method applied with 39 stellar templates. 
\item Based on the dividing lines applied in the BPT diagram of O3HB versus N2HA, 19531 starforming galaxies 
	and 4112 Type-2 AGN are collected with reliable narrow emission lines.
\item Based on the reliable measured parameters of the 19531 starforming galaxies, a strong linear correlation 
	can be confirmed between the narrow H$\alpha$ line luminosity and the continuum luminosity at 5100\AA, 
	with no contributions of AGN activity. The correlation in starforming galaxies can be treated as a 
	standard candle.
\item Based on the reliable measured parameters of the 4112 Type-2 AGN, a strong correlation can be confirmed 
	between the narrow H$\alpha$ line luminosity and the continuum luminosity at 5100\AA, however with quite 
	different intercept from the correlation in the starforming galaxies.
\item Starforming galaxies (Type-2 AGN) with $S/N>30$ can lead to similar correlation between narrow H$\alpha$ 
	line luminosity and continuum luminosity as those of starforming galaxies (Type-2 AGN) with $S/N>30$, 
	leading to few effects of spectral signal-to-noise on the final results.
\item Statistically lower narrow H$\alpha$ line luminosities with significance level higher than 5$\sigma$ can 
	be confirmed for given continuum luminosities in the Type-2 AGN than in the starforming galaxies, lead 
	to apparent clues to support globally negative AGN feedback in Type-2 AGN, even after considering effects 
	of central AGN activity on narrow H$\alpha$ line luminosities, and effects of different distributions 
	of redshift $z$, total stellar mass $M_T$, Petrosian radius $R_{90}$ and inverse concentration parameter 
	$IC$. 
\item Different host galaxy properties can be confirmed with significance levels higher than 
	5$\sigma$ between the starforming galaxies and the Type-2 AGN in the subsamples which have the same 
	distributions of redshift and EW(H$\alpha$). Therefore, besides the different EW(H$\alpha$) properties,
	AGN feedback could be the key role leading to the statistically lower narrow H$\alpha$ line luminosities 
	for given continuum luminosities in the Type-2 AGN than in the starforming galaxies.
\item Comparing properties of Dn(4000) and inverse concentration parameter $IC$ between the 206 starforming 
	galaxies and the 206 Type-2 AGN in the final subsamples with the same distributions of $z$, $M_T$, 
	$R_{90}$ and $IC$ ($IC<0.35$ and devab\_r larger than 0.8), statistically older stellar ages in the 
	Type-2 AGN can provide indirect evidence to support negative AGN feedback in local Type-2 AGN.
\item Under the extreme assumption of no contributions of central AGN activity in the observed narrow 
	H$\alpha$ in Type-2 AGN, the globally lower limited value can be estimated that more than 50\% of 
	narrow H$\alpha$ emissions have been suppressed in the host galaxies of the local SDSS Type-2 AGN 
	due to globally negative AGN feedback.
\end{itemize}

\begin{acknowledgments}
Zhang gratefully acknowledge the anonymous referee for giving us constructive comments and 
suggestions to greatly improve the paper. Zhang gratefully thanks the kind financial support from GuangXi 
University and the kind grant supports from NSFC-12173020 and NSFC-12373014. This manuscript has made use 
of the data from the SDSS projects. The SDSS-III web site is \url{http://www.sdss3.org/}. SDSS-III is managed 
by the Astrophysical Research Consortium for the Participating Institutions of the SDSS-III Collaborations. 
The paper has made use of the LTS\_LINFIT package (\url{www-astro.physics.ox.ac.uk/~mxc/software}) and the 
MPFIT package (\url{https://pages.physics.wisc.edu/~craigm/idl/cmpfit.html}).
\end{acknowledgments}


\label{lastpage}
\end{document}